\pdfoutput=1
\documentclass{article}
\usepackage{spconf,amsmath,graphicx}
\usepackage{booktabs}
\usepackage{hyperref}
\usepackage{float}
\usepackage{diagbox}
\usepackage{multirow}
\usepackage{color}
\usepackage{graphicx}
\usepackage{subfigure}
\usepackage{amssymb}
\usepackage{url}
\usepackage[misc]{ifsym}

\title{Learn2Sing: Target Speaker Singing Voice Synthesis by learning from a Singing Teacher}
\name{Heyang Xue\textsuperscript{1}, Shan Yang\textsuperscript{2}, Yi Lei\textsuperscript{2}, Lei Xie\textsuperscript{1,2}\sthanks{Corresponding author.}, Xiulin Li\textsuperscript{3}}
\address{\textsuperscript{1}Audio, Speech and Language Processing Group (ASLP@NPU),\\
\textsuperscript{1}School of Software,\textsuperscript{2}School of Computer Science, Northwestern Polytechnical University, Xian, China \\
\textsuperscript{3}Databaker (Beijing) Technology Co., Ltd.}
\begin{document}
	%\ninept
	%
	\maketitle
	\begin{abstract}
		
		Singing voice synthesis has been paid rising attention with the rapid development of speech synthesis area. In general, a studio-level singing corpus is usually necessary to produce a natural singing voice from lyrics and music-related transcription. However, such a corpus is difficult to collect since it's hard for many of us to sing like a professional singer. In this paper, we propose an approach -- Learn2Sing that only needs a singing teacher to generate the target speakers' singing voice without their singing voice data. In our approach, a teacher's singing corpus and speech from multiple target speakers are trained in a frame-level auto-regressive acoustic model where singing and speaking share the common speaker embedding and style tag embedding. Meanwhile, since there is no music-related transcription for the target speaker, we use log-scale fundamental frequency~(LF0) as an auxiliary feature as the inputs of the acoustic model for building a unified input representation. In order to enable the target speaker to sing without singing reference audio in the inference stage, a duration model and an LF0 prediction model are also trained. Particularly, we employ domain adversarial training~(DAT) in the acoustic model, which aims to enhance the singing performance of target speakers by disentangling style from acoustic features of singing and speaking data. Our experiments indicate that the proposed approach is capable of synthesizing singing voice for target speaker given only their speech samples.
		
	\end{abstract}
	\begin{keywords}
		text-to-singing, singing voice synthesis, auto-regressive model
	\end{keywords}
	\section{Introduction}
	\label{sec:intro}
	
	\vspace{-0.2cm} 
	Different from text to speech~(TTS) which aims to generate natural speech solely from text~\cite{black2007statistical,ze2013statistical,watts2016hmms,wang2017tacotron}, as another related task, singing voice synthesis~(SVS) mainly focuses on producing singing voice from lyrics and musical scores, where textual lyrics provide linguistic information and musical scores convey pitch and rhythm information. Recently, there have been growing interests in SVS with the fast development of deep learning~\cite{nishimura2016singing,nakamura2019singing,Juntae2018Korean,hono2018recent}. To obtain the satisfactory singing voice of a target speaker, a sizable set of singing recordings with associated lyrics and musical scores are necessary~\cite{gu2020bytesing}. However, collecting and labeling such singing corpus is more difficult and expensive as compared with a speech corpus. A straightforward solution is adapting a multi-speaker singing model with a small amount of the target speaker's singing data~\cite{blaauw2019data}. However, many speakers are not good at singing which makes further difficulty for building singing voice synthesis systems for an arbitrary target speaker.
	
	%since many of us cannot sing well, which limits the applications of singing synthesis. As discussed above, the singing voice can be controlled through lyrics and musical scores, where lyrics or general texts are easy to obtain. Leveraging the availability of speech data, we only need to let the target speaker learn how to produce singing voice through given musical scores.
	
	Alternatively, singing for a new speaker without the speaker's singing data can be achieved by singing voice conversion (SVC) which aims to convert the source singing to the timbre of the target speaker while keeping the linguistic content unchanged. As a recent SVC approach to this end, DurIAN-SC~\cite{zhang2020duriansc} has utilized a unified speech and singing synthesis framework based on the DurIAN TTS architecture proposed earlier~\cite{yu2019durian}. It is capable of generating high-quality target speaker's singing using only his/her normal speech data. However, DurIAN-SC depends on the reference singing audio of the source speaker to generate the singing voice of the target speaker. Mellotron~\cite{valle2019mellotron} is a multi-speaker voice synthesis model based on Tacotron2-GST~\cite{Wang2018Style} that can make a target speaker sing without his singing training data by explicitly conditioning on rhythm and continuous pitch contours. Similar to DurIAN-SC, the pitch contours are extracted from reference audio. This means Mellotron cannot obtain the target speaker's singing voice without reference singing audio.

	%DeepSinger\cite{ren2020deepsinger} was proposed to generate singing samples using data mined from the web instead of professional singing corpus while we focus on training model using standard corpus.
	
	In this paper, we propose \textit{Learn2Sing}, a full singing voice synthesis system for a target speaker without his/her singing data for system building. The proposed model can generate decent singing of the target speaker from lyrics and musical notes without reference singing audios. Inspired by the above approaches especially DurIAN-SC, we learn an SVS system for the target speaker with the help of a singing teacher. Here the \textit{singing teacher} is a singing corpus from a professional singer, while the \textit{target speaker} only has a sizable set of speech and the associated text transcripts. In order to let the target speaker learn the singing patterns from the singing teacher, we propose to treat the speaking and singing data as a whole to train a unified frame-level auto-regressive acoustic model which has a specifically designed encoder-decoder structure.
	
	\vspace{-0.9cm} 
	However, the target speaker has no musical notation such as pitch, and the text representation of speaking and singing data is substantially different, which leads to difficulties to build the above unified model. To obtain a unified input representation for both speaking and singing data, we only keep the common aspects of the two types of data. But the note pitch in the music score is essential for singing. Hence we further use log-scale F0~(LF0) as an auxiliary feature to unify the input representation of the target speaker and the singing teacher.
	
	\vspace{-0.8cm} 
	Besides, there is substantial difference in pronunciation between speaking and singing, which apparently affects the singing performance for the target speaker. In order to disentangle the difference between the two domains, we conduct domain adversarial training (DAT)~\cite{Ganin2014Unsupervised,Ganin2017Domain} in the auto-regressive decoder to obtain style-independent latent features, where a gradient reverse layer~(GRL) followed by a style classifier network is utilized.  And in order to control the domain of generated audio (speak or sing), we inject a binary style tag in the encoder memory to provide style information during training and inference. Note that DAT has been recently used to disentangle speakers in multi-speaker singing synthesis \cite{wujie123}.
	
	\vspace{-0.8cm}
	The proposed Learn2Sing model requires F0 and phoneme duration to generate singing for the target speaker during inference.  So we build a duration model as well as an LF0 prediction model to provide phoneme duration and LF0 during acoustic model inference. Finally, the generated acoustic features are converted to waveform by a Multi-band WaveRNN vocoder~\cite{yu2019durian}.
	
	\vspace{-0.6cm}
	We summarize the contributions as follows:
	\vspace{-2.8cm} 
	\begin{itemize}
		\item We propose a neural sing voice synthesis system -- Learn2Sing, which can generate a singing voice of target speaker without his/her singing data for system training.
		\vspace{-3.3cm}
		\item Different from previous approaches, the proposed approach is capable of generating the target speaker's singing voice without any reference audio.
		\vspace{-3.3cm}
		\item The proposed domain adversarial training strategy can improve the naturalness and expressiveness of singing by learning style-independent latent features that narrow the difference between speaking and singing and enable the style tag to control the singing generation for the target speaker.
	\end{itemize}

	\begin{table*}[htbp]
		\centering
		\caption{Features used for acoustic model (AM), LF0 prediction model (LF0M) and duration model (DM).}\vspace{5pt}
		\begin{tabular}{lccccc}
			\toprule
			\multicolumn{1}{l}{Feature} & \multicolumn{1}{l}{Value/example} & Type  & AM    & LF0M & DM \\
			\midrule
			\multicolumn{1}{l}{\textit{PhonemeID}} & \multicolumn{1}{l}{t, ian} & Categorical     &  \checkmark  & \checkmark    & \checkmark \\	
			\multicolumn{1}{l}{\textit{Pitch}} & \multicolumn{1}{l}{C4, B3} & Categorical     &       & \checkmark     &  \\
			\multicolumn{1}{l}{\textit{Slur}} & \multicolumn{1}{l}{true, false} & Categorical     &       & \checkmark     & \checkmark \\
			\multicolumn{1}{l}{\textit{SpeakerID}} & \multicolumn{1}{l}{0, 1} & Categorical     & \checkmark     &       &  \\
			\multicolumn{1}{l}{\textit{StyleTag}} & \multicolumn{1}{l}{singing, speaking} & Categorical     & \checkmark     &       &  \\
			\multicolumn{1}{l}{\textit{NoteDur}} & \multicolumn{1}{l}{0.6s, 0.3s} & Numerical     &       &       & \checkmark \\
			\multicolumn{1}{l}{\textit{FramePos}} & \multicolumn{1}{l}{0.1, 0.8} & Numerical     & \checkmark     & \checkmark     &  \\
			\multicolumn{1}{l}{\textit{LF0}} & \multicolumn{1}{l}{predicted by LF0M} & Numerical     & \checkmark     &       &  \\
			\midrule
			\multicolumn{3}{r}{Model output} & Mel-spec & LF0   & PhonemeDur \\
			\bottomrule
		\end{tabular}%
		\label{tab:feature}%
	\end{table*}%
	
	\begin{figure}[H]
		
		\begin{minipage}[b]{1.0\linewidth}
			\centering
			{\includegraphics[width=8cm]{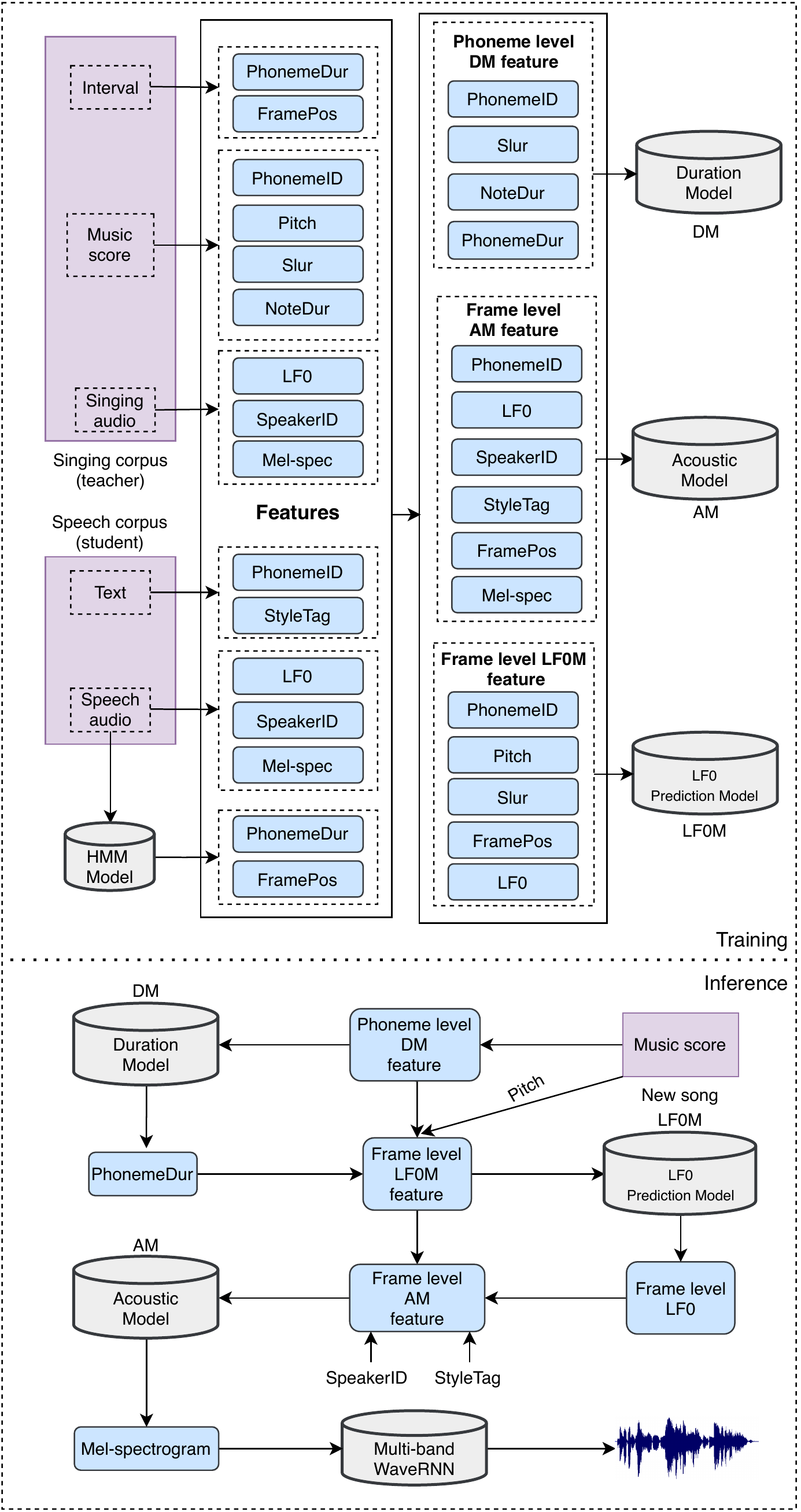}}
			%  \vspace{2.0cm}
		\end{minipage}
		
		\caption{The diagram of the proposed singing voice synthesis approach with training stage and inference stage.}
		\label{fig:framework}
	\end{figure}
	
	\section{System Overview}
	\label{sec:systemoverview}
	
	\vspace{-0.2cm}
	Figure~\ref{fig:framework} illustrates the diagram of our singing voice synthesis approach -- Learn2Sing, which aims to let a target speaker learn to sing from a singing teacher. It mainly consists of two stages -- training and inference (synthesis) and three models to train -- duration model, LF0 prediction model and acoustic model. During the training stage, we use a singing corpus from a professional singer as \textit{teacher} and the speech corpus of the target speaker as \textit{student}. For the singing training data, the phoneme duration comes from the manually labeled interval file which contains the duration interval for each phoneme. An HMM-based force-alignment model~\cite{black2007statistical} is adopted to obtain the phoneme duration for the speech data. The duration model (DM) is trained to predict the phoneme duration using phoneme-level features obtained from the interval and music score. And then, we extract the LF0 directly from the singing and speech audio and combine them with other frame-level features for acoustic model (AM) training. The LF0 prediction model (LF0M) is trained to predict LF0 at frame level from a set of input features. We will introduce the features mentioned above for model training in Section~\ref{sec:features} and the details of the AM, LF0M and DM in Section~\ref{sec:models}.

	During the inference stage, we first extract the phoneme-level duration features from the music scores of a new song and then obtain the phoneme duration from the DM. Right after that, we expand the input features to frame level according to the predicted phoneme duration. The LF0M then takes the frame-level features to predict LF0. The frame-level LF0 is subsequently concatenated with other features as the input of the AM that predicts the mel-spectrogram (Mel-spec). Finally, the Mel-spec is transformed to waveform by Multi-band WaveRNN~\cite{yu2019durian}.

	\section{Features}\label{sec:features}
	
	\vspace{-0.2cm}
	As described in Section~\ref{sec:systemoverview}, we train acoustic, LF0 and duration models to build an SVS system for the target speaker without his/her singing data. Here we describe the features for model training, as summarized in Table~\ref{tab:feature}. The \textit{PhonemeID} and \textit{FramePos} attributes are easily obtained from both speech corpus and singing corpus. \textit{FramePos} is a value in [0, 1], which means the relative position of current frame in a phoneme. \textit{Pitch} is a specific attribute from musical scores to indicate the pitch of singing voice and \textit{Slur} stands for the concept of liaison in music scores.  Same as the typical SVS systems~\cite{gu2020bytesing, Juntae2018Korean}, our duration model (DM) takes [\textit{PhonemeID},~\textit{Slur},~\textit{NoteDur}] as inputs to predict phoneme duration (PhonemeDur) as the duration for each phoneme only depends on the musical scores, where \textit{NoteDur} stands for the note duration calculated from BPM. Beats-per-minute~(BPM) refers to the duration information for each note from the music scores. But for the acoustic model (AM), different from the typical systems~\cite{gu2020bytesing, Juntae2018Korean} that take [\textit{PhonemeID},~\textit{Pitch},~\textit{FramePos}] as inputs, we substitute the \textit{Pitch} attribute with continuous \textit{LF0} predicted from the LF0 model because we build a unified AM for both speaking and singing, which desires a unified input as well. Moreover, we include two extra tags in AM --\textit{SpeakerID} and \textit{StyleTag}, so that we can explicitly distinguish between singing and speaking during training. But in inference, we just simply set \textit{StyleTag} to singing for the target speaker so that singing voice can be generated for the target speaker. The LF0 prediction model takes [\textit{PhonemeID},~\textit{Pitch},~\textit{Slur},~\textit{FramePos}] as inputs to estimate frame-level LF0 value, where those features are from musical scores and interval.
	
	\section{Models}\label{sec:models}
	\vspace{-0.2cm}
	\subsection{Acoustic model}
	\label{sec:am}
	
	Figure~\ref{fig:acoustic_model} is the encoder-decoder architecture of our frame-level acoustic model showing how we generate singing mel-spectrogram for target speaker who has no singing data for model training. As described earlier, during the training stage, we train a unified AM for both speaking and singing, which shares common speaker embedding and style tag embedding. \textit{PhonemeID}, \textit{FramePos} and \textit{SpeakerID} are first encoded by embedding/dense layer and concatenated together to feed to the CBHG-structured encoder. \textit{StyleTag} and \textit{LF0} are encoded by an embedding layer and a dense layer respectively and further concatenated with the encoder output before going through the decoder RNNs. The DecoderRNN takes the current time step encoder output and previous time step mel-spectrogram which is passed through a PreNet as inputs to predict the current time step mel-spectrogram. The output mel-spectrogram from the DecoderRNN is passed through a PostNet to predict the residuals. The module in the dotted box is the domain adversarial training module which will be described in Section~\ref{subsub:DAT}. During inference,~\textit{FramePos} and \textit{LF0} are predicted from the duration and LF0 prediction models respectively. Alternatively, \textit{FramePos} can be obtained from \textit{PhonemeDur} according to the interval file if it is available.
	
	\subsubsection{Domain adversarial training}\label{subsub:DAT}
	
\begin{figure}[t]
		
		\begin{minipage}[t]{1.0\linewidth}
			
			\leftline{\includegraphics[width=8cm]{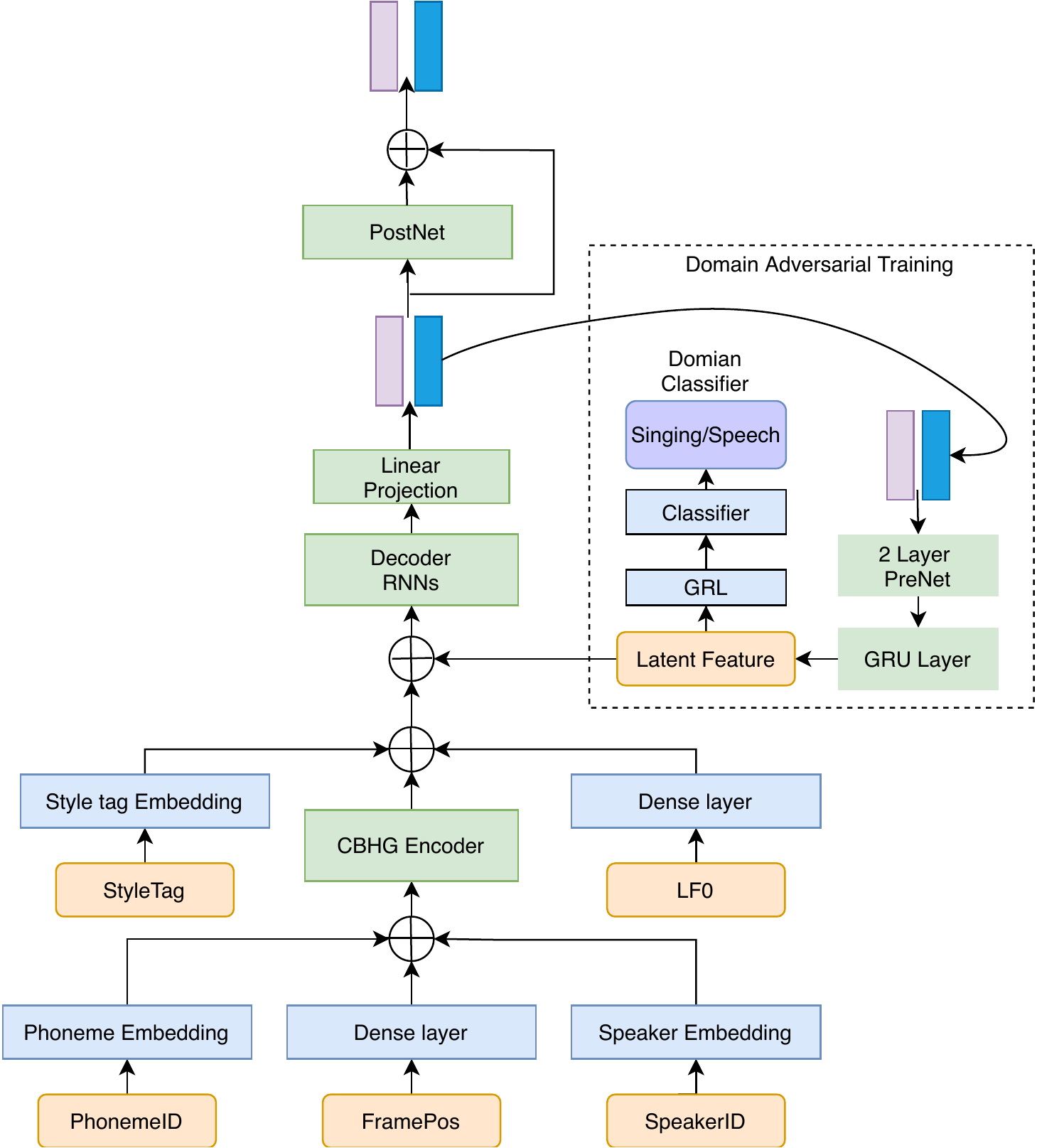}}
			%  \vspace{2.0cm}
		\end{minipage}
		
		\caption{The diagram of the acoustic model with the domain adversarial training module.}
		\label{fig:acoustic_model}
	\end{figure}

	Although speaking and singing are from the same human articulation system, there are still substantial differences in acoustic outcomes. We find that if we set the \textit{StyleTag} to singing for the target speaker during inference with the unified AM and unified feature inputs introduced earlier, the synthesized singing sometimes is still similar to speech with shortened phoneme duration and lack of expressivity (see Figure~\ref{fig:compare_mel} in Section~\ref{sec:exp}). In order to further disentangle the style (speaking vs. singing) information in the acoustic space learned by the AM and make the style of generated voice determined by \textit{StayTag}, we adopt a domain adversarial training (DAT) module to learn style-invariant latent feature, as shown in the dotted box in Figure~\ref{fig:acoustic_model}. Specifically, we inject a GRU layer after PreNet and employ a style classifier with a gradient reversal layer (GRL) on the output of the GRU layer. The latent feature is supposed to be style-invariant according to the theory of DAT~\cite{Ganin2014Unsupervised}. We concatenate the latent feature with the \textit{StyleTag} embedding so that we can enhance the function of \textit{StyleTag} in generating the singing voice of the target speaker.
	
	We minimize the cross entropy loss for the style classifier:
	\begin{equation}
	\mathcal{L}_{\mathrm{adv}}=\sum_{i}\mathcal{L}_{ce}(S(G(z_{i})), s_{i}) \label{eq:adv}
	\end{equation}
	where $S$ is the style classifier and $G$ represents the GRU layer. $z_i$ denotes the latent feature after the PreNet for the $ith$ frame while $s_i$ denotes the \textit{StyleTag} for the $i$th frame.

	\subsubsection{Loss Function}
	
	The AM takes frame-level features as inputs and generates mel-spectrogram for the target speaker. The reconstruction loss of the AM is the mean-square error~(MSE) of mel-spectrogram before and after the PostNet as well as the $l2$ regularization loss, defined as
	
	\begin{equation}
	\mathcal{L}_{\mathrm{rcon}}\!=\!\mathcal{L}_{\mathrm{mse}}({M},\! \hat{M}) \! + \mathcal{L}_{l2}
	\end{equation}
	where $M$ and $\hat{M}$ denote the ground-truth and predicted mel-spectrogram respectively.
	
	Adding the adversarial loss in Eq.~(\ref{eq:adv}), the final loss for the AM becomes:
	\begin{equation}
	\mathcal{L}_{total}=\mathcal{L}_{\mathrm{rcon}}+\lambda* \mathcal{L}_{\mathrm{adv}}
	\end{equation}
	where $\lambda$ is a tunable weight.

	\subsection{Duration and LF0 prediction model}
	
	When reference singing audio is available in the inference stage, we can extract LF0 from the reference audio, and the phoneme duration can be obtained from the interval file. In this case, we can perform singing voice conversion and we do not need duration or LF0 prediction. In order to make the target speaker sing only from the music scores, i.e. achieving a fully singing voice synthesis system, we then need to train individual components to predict duration and LF0.

	The duration model (DM) predicts phoneme duration and expand the input to frame level for the acoustic and LF0 models during the inference phase. In typical TTS systems, the duration model usually predicts frames of phoneme directly by minimizing the mean squared error~(MSE) between the ground truth duration and the predicted value. Instead, we construct a Mixture Density Network~(MDN)~\cite{bishop1994mixture} to model the conditional probability distribution to parameterize a Gaussian mixture model~(GMM). The goal of this MDN model is to encode the music score information and to predict the parameters of GMMs. %The general expression is as follows:
	
	We minimize the negative log-likelihood~(NLL) loss as:
	\begin{equation}
	\mathcal{L} = -\frac{1}{N} \sum_{i=1}^{N} \log \sum_{k=1}^{K} \pi_{k} {N}\left({y}^{target}_{i} | \mu_{i k}({x}). \sigma_{i k}^{2}({x})\right),
	\label{eq2}
	\end{equation}
	where $x$ is the encoder output, $ {y}^{target}$ is the target frames, K is the number of mixture, N is the total frames, ${\pi_{k}(x)}$ is the mixing coefficients, and $\mu_{k}(x)$ and $\sigma_{k}^{2}(x)$ are the mean and variance vectors of the Gaussian component density respectively.
	During the training stage, $ {y}^{target}$ is set to frames which obtained from the interval files. During inference, duration is generated by sampling from the corresponding Gaussian component.

	For the LF0 prediction model, we deploy a frame-level MDN which has the same structure as the DM. And in the inference phase, we first expand the input features to frame level using the phoneme duration predicted by the duration model. As described in Section~\ref{sec:am}, the predicted LF0 values are used as part of the inputs for the acoustic model.

	\section{Experiments}
	\label{sec:exp}

	\subsection{Basic setups}
	In our experiments, we use the internal singing data as the professional singing teacher. It consists of 5 hours of singing data, which contains about 100 Chinese songs recorded by a female singer with lyrics, corresponding musical scores and interval files. We treat two female TTS corpora as students: one from the open-source TTS corpus Databaker-DB1\footnote{Available at: \url{www.data-baker.com/open_source.html}} (as Student-1) and another is an internal TTS corpus (as Student-2). To balance the amount of singing and speaking data, we also use 5 hours of training data from each TTS corpus. The goal of this paper is to teach the two students to know how to sing from the singing teacher.
	
	We down-sample both the singing and TTS audio to 24kHz from 48kHz and extract 80-band mel-scale spectrogram as the decoder targets. We adopt Reaper~\cite{talkin2015reaper} to extract continuous F0 from waveform and convert it to log-scale (LF0) to control the acoustic model. We train three speaker-dependent Multi-band WaveRNNs~\cite{yu2019durian} to reconstruct waveform from generated mel-spectrogram. We reserve three complete songs and 50 pieces of songs with related musical scores from the singing corpus for evaluation, and the rest is used for training and validation. To evaluate the proposed model, we conduct mean opinion score (MOS) on the generated singing voice from different methods, and 20 listeners are attending this subjective evaluation.

	\subsection{Model parameters}
	For the acoustic model, we adopt a CBHG~\cite{wang2017tacotron} module as encoder, which contains 16 sets of 1-D convolutional filters, 2 1-D convolution layers, 4 layers of highway networks and a bidirectional GRU. The number of channels in the encoder convolutions is 256 and the kernel size of all convolutional layers is 3. 
	
	For the auto-regressive decoder, we follow the design of Tacotron2~\cite{shen2018natural} to produce mel-spectrogram. Noted that we conduct frame-to-frame mapping during training and inference, so the decoder doesn't contain the attention mechanism to align encoder outputs and acoustic targets. In detail, in each time step $t$, we utilize previous hidden states of decoder RNN and the $t$th frame of encoder outputs as context vector and to produce $t$th acoustic frame. As for the domain adversarial module, we inject a GRU layer after the decoder PreNet to provide style-independent features.  The style classifier in the DAT module is simply composed of two dense layers to predict style categories. We set the ratio of the style classifier loss  $\lambda$ to 0.001. Besides, we also adopt another CBHG module as PostNet~\cite{shen2018natural} to obtain final mel-spectrograms.
	
	For the duration and the LF0 prediction model, we both use a CBHG encoder followed by three dense layers to predict the GMM parameters. And the mixture number of each model is both set to 8.
	
	\setlength{\tabcolsep}{6mm}{	
		\begin{table}[H]
			\centering
			\caption{Objective criteria of duration and LF0 prediction models}
			\vspace{5pt}
			\begin{tabular}{rl}
				\toprule
				Criterion       &   Value         \\ \midrule
				Duration-Accuracy  &   85.43\% \\
				LF0-PCC       &   0.8835  \\
				LF0-RMSE       &   25.97   \\ \bottomrule
			\end{tabular}
			\label{duration_lf0}
		\end{table}
	}
	\subsection{Evaluation of LF0 and duration model}
	
	Since our method needs individual models to provide LF0 and phoneme duration information during inference, we first evaluate the performance of the LF0 and duration model, as shown in Table.~\ref{duration_lf0}. For the duration model, we calculate the Duration-Accuracy defined as:
	\begin{equation}
	Accuracy=1-\left ( \frac{\sum_{N}^{i}\left | predict_{i}-real_{i}\right |}{\sum_{N}^{i}max(predict_{i}, real_{i})}\right ) ,
	\end{equation}
	where $N$ is the total number of all phonemes in test utterances, $predict_{i}$ is the predicted frames of the $i$th phoneme and $real_{i}$ is the ground truth frames of the $i$th phoneme.
	
	As for the performance of the LF0 model, we evaluate the Root Mean Squared Error~(RMSE) and the Pearson Correlation Coefficient (PCC). Since the predicted LF0 and phoneme duration are eventually used to generate mel-spectrograms in the following acoustic model, we will analyze their effects on the acoustic prediction in more convincingly subjective listening tests.

	\subsection{Evaluation of acoustic model}
	
	\setlength{\tabcolsep}{1.5mm}{
		
		\begin{table*}[htbp]
			\centering
			\caption{The results of subjective evaluation in terms of MOS with confidence interval: 95\% }
			\begin{tabular}{clllccc}
				\toprule
				\multicolumn{1}{c }{Target speaker} &	\multicolumn{2}{l}{Test condition} & \multicolumn{1}{l}{Model}  & \multicolumn{1}{c}{Naturalness} & \multicolumn{1}{c}{Expressiveness} & \multicolumn{1}{c}{Similarity} \\
				\midrule
				\multirow{4}[0]{*}{Student-1} &		\multicolumn{2}{l}{\multirow{2}[0]{*}{Real \textit{PhonemeDur} \& \textit{LF0}}} &  baseline   & 2.69$\pm$0.045& 2.68$\pm$0.051   & 3.03$\pm$0.077 \\
				&	\multicolumn{2}{l}{\multirow{2}[6]{*}{Predicted \textit{PhonemeDur} \& \textit{LF0}}} & baseline + DAT     &3.41$\pm$0.047 & 3.44$\pm$0.043 & 3.29$\pm$0.080  \\
				\multicolumn{1}{c}{} &	\multicolumn{2}{c}{} & baseline    & 2.70$\pm$0.042  & 2.66$\pm$0.050& 3.03$\pm$0.078 \\
				\multicolumn{1}{c}{} &	\multicolumn{2}{c}{} &  baseline + DAT  & 3.24$\pm$0.048 & 3.24$\pm$0.046   & 3.20$\pm$0.088\\
				\midrule
				\multirow{4}[0]{*}{Student-2} &				\multicolumn{2}{l}{\multirow{2}[0]{*}{Real \textit{PhonemeDur} \& \textit{LF0}}} & baseline  &2.48$\pm$0.047 &2.54$\pm$0.056  & 3.02$\pm$0.080 \\
				
				&	\multicolumn{2}{l}{\multirow{2}[6]{*}{Predicted \textit{PhonemeDur} \& \textit{LF0}}} &  baseline + DAT  & 3.48$\pm$0.047   & 3.50$\pm$0.043   & 3.29$\pm$0.083 \\
				\multicolumn{1}{c}{} &		\multicolumn{2}{c}{} &  baseline & 2.52$\pm$0.050  &2.60$\pm$0.057  & 3.01$\pm$0.082 \\
				\multicolumn{1}{c}{} &		\multicolumn{2}{c}{} &  baseline + DAT & 3.23$\pm$0.052   & 3.25 $\pm$0.047  & 3.31$\pm$0.082 \\
				\midrule
				\multirow{2}[0]{*}{Teacher} &		\multicolumn{2}{l}{\multirow{1}[0]{*}{Synthesized singing}} 	&	-&  3.90$\pm$0.036   & 3.89$\pm$0.029   & 3.69$\pm$0.097 \\
				& 	\multicolumn{2}{l}{\multirow{1}[0]{*}{Original singing}}  & -& 4.66$\pm$0.031   & 4.67$\pm$0.031   & - \\
				\bottomrule
			\end{tabular}%
			\label{tab:addlabel}%
			\vspace{-0.3cm}
		\end{table*}%
		
	}
	
	Note that we can not evaluate the acoustic model in an objective criterion since there is no ground truth singing audio for the target speakers. we compare the proposed models with mean opinion score (MOS) in terms of naturalness, expressiveness and similarity, where naturalness is considered as the accuracy of pronunciation and singing performance, while expressiveness is considered as singing skill and rhythm. The above two aspects aim to measure the singing skills of the students learned from the singing teacher. Besides, we conduct a similarity test between the synthesized singing and the real audio from the target speaker. Since the singing audio is unavailable for the two students, we just let the listeners to compare the synthesized singing with their original speaking recordings.
	
	To evaluate the proposed Learn2Sing, we firstly build a singing voice synthesis system of the singing teacher to show the ceiling of our singing performance. The skeleton of this system is the same as the Learn2Sing system without the DAT module, while we use ground-truth duration to synthesize the test songs. As shown in Table.~\ref{tab:addlabel}, the performance of the singing teacher achieves MOS value about 3.9 in both naturalness and expressiveness, which indicates that our basic framework has the ability to conduct singing voice synthesis well.
	
	We then evaluate the performance of baseline system without DAT module using ground-truth duration and LF0. For both speaking students, although the MOS values are significantly lower than the top-line singing teacher, we find the proposed model can generate their singing voice, which achieves our goal to make target speakers sing without their singing data. Based on the baseline acoustic model, we also compare its performance with predicted LF0 and phoneme duration. The subjective results show that there are no significant differences when we use predicted auxiliary features. This result means that the MDN based LF0 and duration prediction model can produce credible outputs for the baseline acoustic model.
	
	As discussed above, there are substantial differences in acoustic outcomes of speaking and singing, so we further evaluate the performance of the baseline system with DAT module. As shown in Table.~\ref{tab:addlabel}, the baseline + DAT significantly outperforms the baseline model for both speakers in all aspects, which achieves MOS values over 3.4 for both naturalness and expressiveness with real phoneme duration and LF0. As for the predicted phoneme duration and LF0, DAT also helps a lot with big MOS improvement. Meanwhile, speaker similarity also improved substantially.  In order to unveil why the listeners prefer the baseline + DAT, we visualize several testing examples of two singing segments from the two students in Figure.~\ref{fig:compare_mel}. Compared with the baseline model, we can find that the baseline + DAT model can generate better vocal performance with appropriate timing and spectrogram with finer details, which leads to significant improvement in singing naturalness and expressiveness. We suggest the readers listen to the examples from: \url{https://welkinyang.github.io/learn2sing/}
	
	\begin{figure}[ht]
		\begin{minipage}[b]{1.0\linewidth}
			\centering
			{\includegraphics[width=8.5cm]{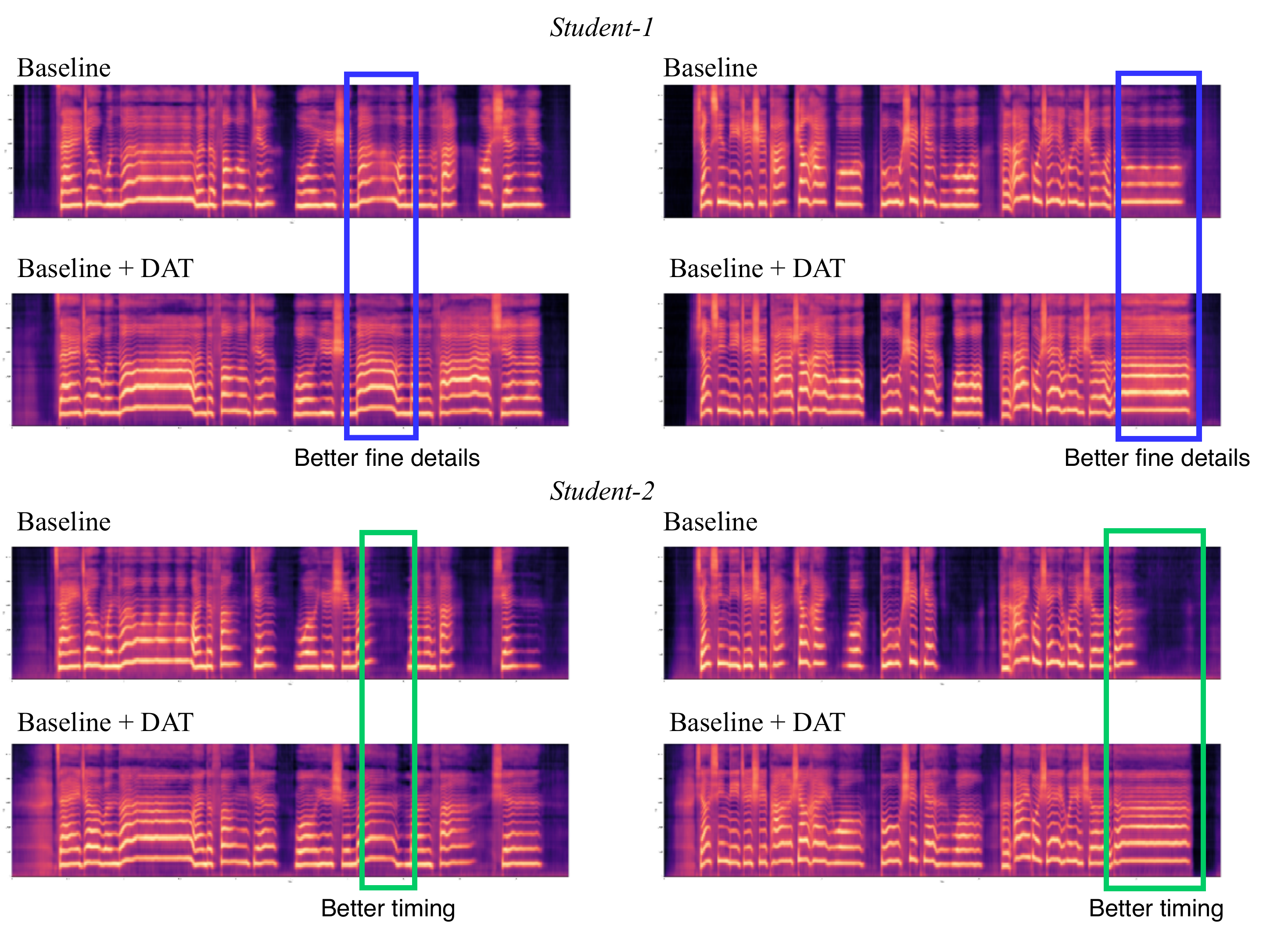}}
			%  \vspace{2.0cm}
		\end{minipage}
		\vspace{-0.5cm}
		\caption{Mel-spectrogram of testing samples for different models.}
		\label{fig:compare_mel}
	\end{figure}
	
	\vspace{-1cm}
	\section{Conclusions}
	
	In this paper, we propose \textit{Learn2Sing}, which aims at producing a singing voice of target speakers without their singing data for model training. To achieve this goal, we build a unified frame-level acoustic model to model both speaking and singing. Considering the pitch in music score plays a decisive role in singing voice synthesis, we utilize an auxiliary LF0 feature extracted from singing and speaking audio to provide such information. In this way, we can organize the same linguistic representations for both text and lyrics to build the above unified model. In order to disentangle style information in speaking and singing, we further adopt domain adversarial training to learn style-independent features to improve the model performance. The experimental results show that the proposed Learn2Sing can teach the target speaking student to sing from a singing teacher, and the DAT module can further improve the singing performance of target students.
	
	%\section{ACKNOWLEDGMENTS}
	%The author would like to thanks


	\bibliographystyle{IEEEbib}
	
\end{document}